# Bayer-type Vis-NIR Routing via Inverse Design for Submicron-pixel Image Sensing Chip

*Xianguang Yang\*, Shijie Xiong, Fangchang Tan, Zhitao Lin, Yanjun Bao, Long Wen, Qin Chen, and Baojun Li*

Guangdong Provincial Key Laboratory of Nanophotonic Manipulation, Institute of Nanophotonics, College of Physics & Optoelectronic Engineering, Jinan University, Guangzhou 511443, China

\*Email: xianguang@jnu.edu.cn

**Abstract**

With the advent of high-precision nanoscale lithography technology, high-resolution image sensing has experienced rapid development in recent years. Currently, mainstream commercial image sensors predominantly utilize Bayer array color filters to implement RGB colorful imaging strategies. However, as pixel sizes transition into the submicron dimensions, traditional dye filters used in image sensors have long been hampered by limited optical efficiency, suboptimal signal-to-noise ratios, and significant difficulties in miniaturization. In this work, a novel 4-channel RGB-IR color router for image sensing, distinct from the traditional absorption-transmission mechanisms, was proposed through inverse design methodologies. Utilizing genetic algorithms and DCGAN models, approximately 20,000 random color routing structures were generated and trained. From these, an optimized spectral splitting structure with a minimal periodic size of 1.6 μm × 1.6 μm was identified. This structure achieves peak optical efficiencies 1.7 times greater than those of dye filters, while also offering superior color imaging quality and signal intensity. This innovative design approach, leveraging deep learning integration, demonstrates an on-chip strategy for color realization in 4-channel image sensors, and holds significant promise for enhancing the development of next-generation high-performance image sensing chip systems.

**Introduction**

Owing to the advancements in various image recognition technologies and artificial intelligence, research related to image sensors has gained significant attentions in recent years.[1,2] By leveraging optical elements such as spectral tuning, polarization, and phase manipulation, image sensors can achieve advanced optical imaging quality across diverse application scenarios, solidifying their status as essential components for extending human visual capabilities.[3-5] Due to the developments in etching processes, nanoscale lithography, and microelectronics, image sensors have been progressing towards submicron-level pixel sizes and large scale arrays to enhance image resolution.[6] Presently, commercial image color solutions primarily employ RGB dye filters, which operate based on the selective transmission and absorption of organic dyes.[7,8] This operational mechanism results in a substantial waste of available light flux, imposing a physical limitation on optical efficiency. In contemporary color imaging systems, incident light typically undergoes segmentation within each periodic unit cell, as exemplified by the R-G-G-B configuration of Bayer arrays.[9,10] As depicted in **Figure 1**a, at each pixel unit, dye filters selectively sieve out light flux not pertaining to the desired pixel band, resulting in an only quarter of the light flux each pixel can effectively utilize, while the residual light flux is regrettably wasted. Thus, even with dye filters boasting a transmittance of 100%, the average maximum optical efficiency merely amounts to 25%. It is noteworthy that the essence of color management in image sensors encompasses the process of spectral truncation and spatial guidance of broadband incident light. Unlike the dye filtration mechanism, several other spectral splitting schemes also hold certain potential applications. Devices such as diffractive gratings,[11] dielectric nano-antennas,[12] metallic surface plasmon nanostructures,[13] and metasurfaces can modulate the wavefront of the light beam through varying degrees of phase modulation techniques.

[14] Ultimately, altering the refractive index distribution in three-dimensional space enables the acquisition of spatial dispersion of the light beam. More specifically, a diffraction-based spectral splitting structure utilizing nanodielectric materials has been demonstrated to achieve the primary color separation of 430 nm (B), 520 nm (G), and 635 nm (R) on an image sensor with a pixel size of 1.4 μm × 1.4 μm by inducing phase delays through SiN nanopillars with varying lateral dimensions.[15] Another metalens with a dual-layer nanopillar array offers enhanced design flexibility and phase modulation capabilities. This structure can achieve spectral splitting at 1180 and 1680 nm with optical efficiencies of 38% and 52%, respectively. When operating with a dual-layer metalens of 500 μm in diameter and a focal distance of 3 μm.[16] More strikingly, an algorithmically optimized random color routing structure achieving spectral splitting at 650 nm, 540 nm, and 450 nm on a Bayer array with a minimal period size of 2 μm × 2 μm, boasting optical efficiencies as high as 58%, 59%, and 49%, respectively.[17] Compared to other methods, our novel method achieved on-chip spectral splitting structure exhibits superior optical efficiency while being less affected by size constraints, thus holding tremendous potential for the development of high-performance on-chip integrated image sensors.

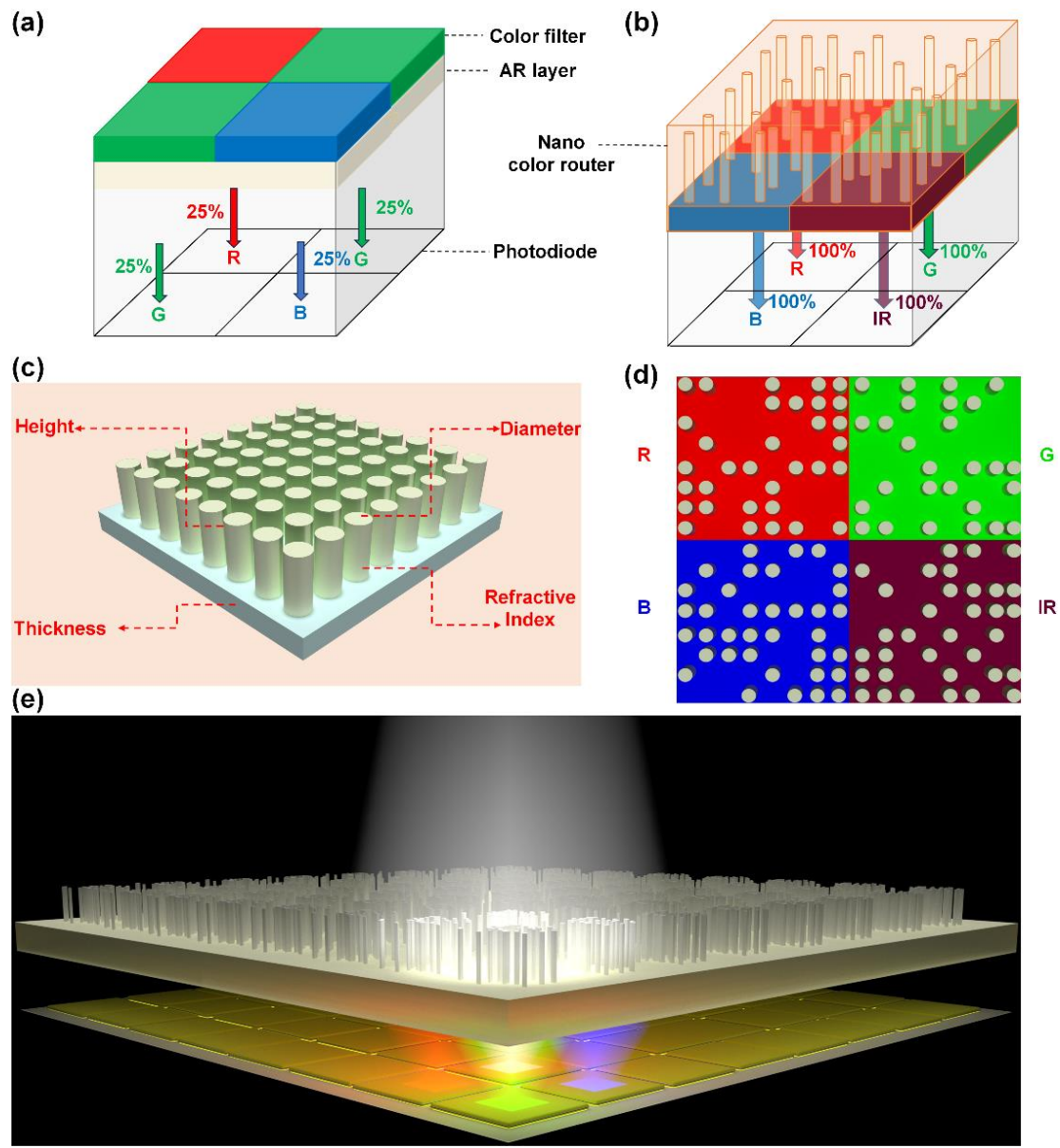

**Figure 1.** Diagrammatic representations of spectral splitting strategies and color routing structures for image sensors. (a) Schematic diagram of RGB three-channel color imaging with dye filter. (b) Color routing of RGB-IR Bayer array with four channels. (c) Schematic diagram of the nano-pillar array utilized for color routing design, including parameters such as pillar height, diameter, refractive index, and substrate thickness. (d) Array structures of RGB-IR color routing designed by genetic algorithms, with dimensions of 1.6 μm × 1.6 μm. (e) Schematic diagram of the four-channel color routing obtained by inverse design.

**Results and discussion**

In this work, we demonstrated a color routing structure that has been carefully optimized through inverse design, consisting of a single-layer random nano-pillar array. As illustrated in **Figure 1**b, we have innovatively extended the three-color channels into the near-infrared spectrum, evolving the traditional RGGB Bayer array into an RGB-IR four-channel spectral design to unlock further applications within the near-infrared range. Through incorporating multi-objective optimization algorithms into deep learning techniques to assist the inverse design process, this spectral strategy effectively circumvents the inherent physical limitation of light flux wastage within dye filter mechanisms.[18-20] Furthermore, by adjusting the arrangement of nano-pillars to achieve wavelength-dependent spatial phase distributions, it enables directional guidance and manipulation of light in both the spatial and frequency domains. Theoretically, our method can achieve 100% routing efficiency under specific configurations, resulting in a nearly four-fold enhancement in optical efficiency. Specifically, when employing the Finite-Difference Time-Domain (FDTD) method for simulating and designing nano-pillar arrays,[21] it primarily involves physical parameters such as the diameter $d$, height $h$, refractive index $n$, and the thickness $t$ of the substrate, where monitors are placed. As illustrated in **Figure 1**c, we conducted simulation tests and correlation coefficient analyses on various structural parameters of the nano-pillar array, considering a parameter variation range of 10%. Details can be found in **Figure S1**. Based on refractive index and fabrication tolerance in reported literature,[22] we opted for high-refractive-index $TiO_2$ nano-pillars as the building block of spectral routing design, while $SiO_2$ serving as the substrate material supporting these nano-pillars. At the pixel scale, we designed each pixel to 800 nm in size and established a minimum period of 1.6 μm × 1.6 μm for the 4-channel configuration. As depicted in **Figure 1**d, the spatial arrangement of four pixels utilized single 100 nm diameter nano-pillar, with a structured layout of 16 × 16 determined through multi-objective genetic algorithms, randomly positioned within their spatial coordinates. **Figure 1**e shows the on-chip integration of a single-layer routing structure consisting multiple pixels periods achieved through inverse design, facilitating a four-channel spectral routing process for chip-scale image sensing. It shows that the broadband incident light strikes an array of pillars on a substrate, after passing through this single-layer routing structure, RGB-IR light beams strike pixels of the same color on a photodiode chip surface. Which indicating that our design enabled on-chip light field manipulation in both spatial and frequency domains. This strategic design effectively mitigates the inherent inefficiencies of light flux utilization posed by dye filter mechanisms, which may potentially advance the on-chip integration of single-layer sub-micron pixel spectral routing structures with current image sensors.[23,24] Such integration holds immense promise across diverse fields including smartphone technology, autonomous vehicle vision perception systems, low-light near-infrared night vision, and image navigation.

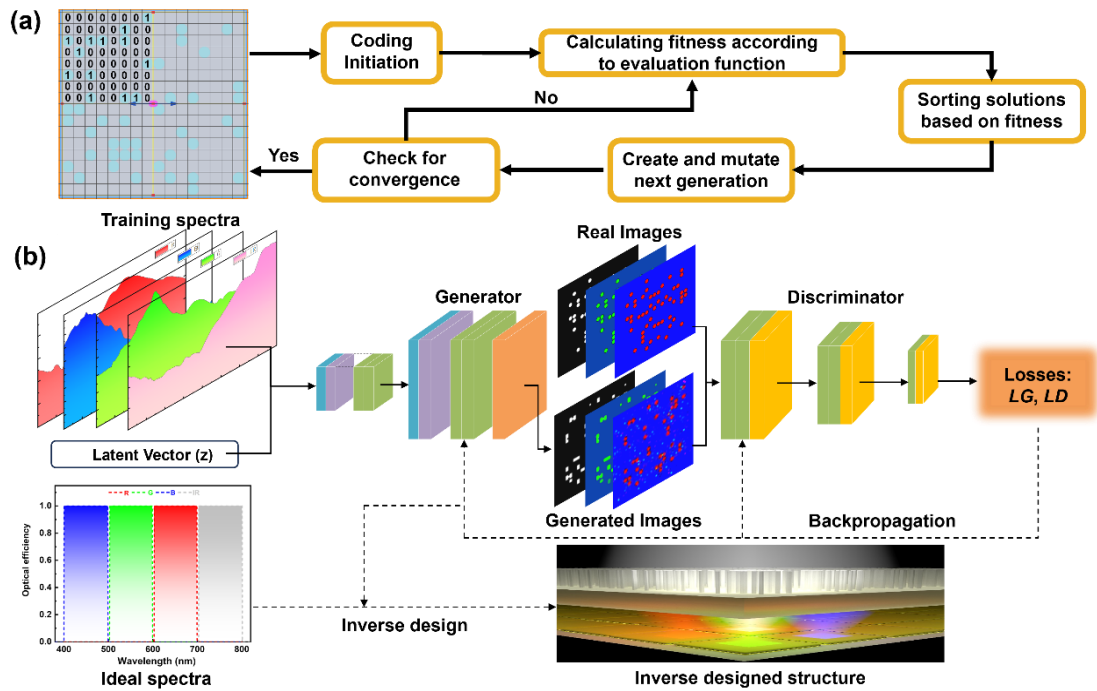

**Figure 2.** Genetic algorithm optimization process and schematic diagram of deep learning inverse design color routing structures. (a) Design flowchart of multi-objective genetic algorithm NSGA-II encoding and optimizing color routing structures, where the value 0 and 1 indicates without and with nano-pillar positioned. (b) Schematic diagram depicting the inverse design process in which routing spectra and images of routing structures, optimized via the genetic algorithm, undergo

generative adversarial training using the DCGAN model.

During the optimization process by inverse design, it is important to consider the spectral response corresponding to the RGB-IR channels for each routing structure. Consequently, the optimization algorithm must concurrently optimize the light flux across all four channels, striving for solutions that maximize the light flux for each channel. Esteemed numerical optimization algorithms employed for this purpose include Particle Swarm Optimization (PSO),[25] Genetic Algorithms,[26-28] Simulated Annealing,[29] Ant Colony Optimization,[30] and Gradient Descent.[31] Among these, the NSGA-II multi-objective optimization algorithm is particularly well-suited for the problem of spectral routing optimization.[32] In the context of spectral routing, the primary optimization objectives of the algorithm are the optical efficiencies of the four channels. By continually generating spectral routing structures as the genetic population, the algorithm iteratively evolves and updates the population through crossover and mutation. This process ultimately yields a routing structure with optimized spectral splitting performance. As illustrated in **Figure 2**a, the 16 ×16 grid cells of the color routing structure are initially encoded in the FDTD simulation using the NSGA-II algorithm. Where 0 indicates the absence of a nanopillar at a given grid position, and 1 signifies the presence of a nanopillar. During each initialization of the genetic algorithm, 40 routing structures are generated. The optical flux across the channels is then used to evaluate the fitness of each routing structure. Following a rapid non-dominated sorting process, superior individuals within the population undergo crossover and mutation operations, yielding a new generation for the subsequent iteration. This iterative process continues, progressively refining the routing structures toward optimal spectral splitting performance. Although employing a random 16 ×16 nanopillar arrangement provides significant design freedom in the search space, the strategy of iteratively generating new routing structures through a genetic algorithm is inherently "exhaustive". Relying only on this method for optimization can result in substantial time and resource consumption, potentially leading to inefficiencies and wastage. Therefore, in the optimization process, we uniquely incorporated an inverse design strategy rooted in deep learning.[33,34] At the specific design level, we initially utilized NSGA-II for preliminary optimization, resulting in approximately 20,000 randomly generated routing structures. Subsequently, these routing structures underwent weighted spectral transformation to form a training spectral dataset. Ideally, the target spectral response corresponds primarily to the red, green, blue, and near-infrared channels at wavebands of 600-700 nm, 500-600 nm, 400-500 nm, and 700-800 nm, respectively, as depicted in **Figure 2**b. Then, a Deep Convolutional Generative Adversarial Network (DCGAN) was introduced for inverse design.[35,36] Given DCGAN's notable advantages in generating images from textual descriptions, it proved highly effective in predicting routing structures from spectral response in image sensing applications. While obtaining the spectral response profiles of random routing structure, we concurrently translated the arrangement codes of each routing structure into MATLAB to generate images at a resolution of 64 × 64 pixels. Additionally, we mapped the structural parameters of nano-pillars and substrates to color channels, thereby transforming the three-dimensional structural parameters into a two-dimensional image.[37] Specifically, the blue channel was mapped to the thickness of the substrate, the red channel to the refractive index of the nano-pillars, and the green channel to the height of the nano-pillars. Then, extensive simulations were conducted under parameter settings where the substrate thickness was 2 μm, the nano-pillar refractive index was 2.4, and the nano-pillar height was 600 nm, creating a dataset of images corresponding to the spectral dataset. In the DCGAN model, there are primarily two modules: the generator and the discriminator. The generator receives training spectral datasets and concatenates them with a random latent vector, transforming them into textual information identifiable in dimensionality.[38-40] Through a series of internal transformations using feature maps,[41] the generator outputs color images at the resolution of 64 × 64 pixels. Simultaneously, the discriminator processes both pre-constructed real image datasets and generated color image datasets. During iterative refinement, the generator and discriminator exchange loss values, *LG* and *LD*, through backpropagation of errors. After numerous training iterations, the model acquires the capability to predict routing structures inversely from spectral response inputs.[42] Consequently, when ideal spectral response profiles are fed into the trained model, it effectively predicts routing structures in reverse, thereby fulfilling the inverse design objectives.

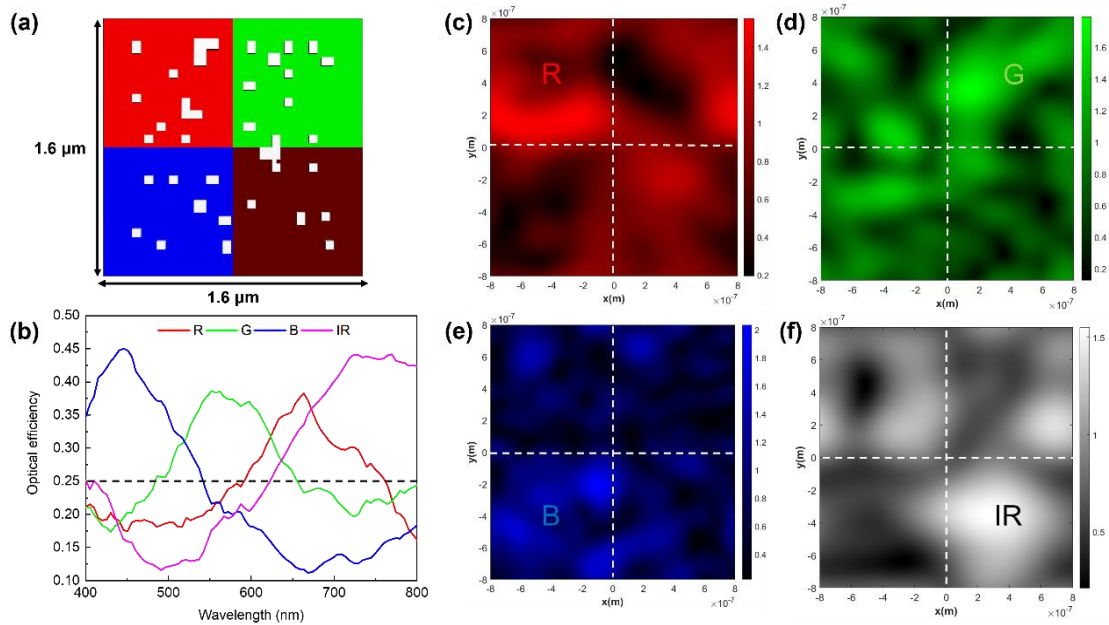

**Figure 3.** Splitting spectra and electric field distribution for single-period inversely designed color routing structure. (a) The color routing structure, optimized through inverse design, features randomly generated $TiO_2$ nano-pillars (white areas), with each pixel exhibiting spatial sizes of 800 nm. (b) The splitting spectra of optimized structure delineate the blue channel centered at 400-500 nm, the green channel at 500-600 nm, the red channel at 600-700 nm, and the near-infrared channel at 700-800 nm. Dashed lines denote the maximum optical efficiency of 25% for the four-channel dye filtering mechanism. (c-f) Spatial electric field distributions for red, green, blue, and near-infrared pixels, respectively, showcasing localized electric field patterns at 650 nm, 550 nm, 450 nm, and 750 nm wavelengths.

It is noteworthy that while employing DCGAN for inverse design, we also evaluated the model's capacity to predict spectral response from routing structures. After rigorous training, it was found that the forward prediction accuracy of DCGAN for the weighted new spectra reached an impressive 90%-95%, as detailed in **Figure S2**. This indicates that the trained DCGAN has acquired a substantial learning capability for "spectrum-to-structure" relationships. Furthermore, as the number of training iterations increases during the inverse design process, the loss values of both the generator and the discriminator exhibit a trend of progressively widening divergence. This trend is positively correlated to a certain extent with the quality of the images generated by the generator. After more than 500,000 iterations of the training process, the optimized inverse design of the color routing structure was achieved, as depicted in **Figure 3**a. It is noteworthy that, due to the low resolution of the images used in the inverse design, the process of importing the generated images into the FDTD for reconstruction involves a grayscale pixel processing algorithm that trims circular pixel boundaries into rectangular shapes.[43] Consequently, the final inverse-designed routing structure exhibits a rectangular morphology. As shown in **Figure 3**b, the inverse-designed routing structure exhibits an average optical efficiency across the four channels that is approximately 1.7 times greater than the 25% maximum optical efficiency of conventional dye filters. Notably, the blue and near-infrared channels achieve optical efficiencies exceeding 45%, while the green and red channels attain approximately 40%. This indicates that our inverse-designed routing structure offers a distinct advantage in terms of optical efficiency. Furthermore, **Figures 3**c to **3**e display the simulated electric field distributions at wavelengths of 650 nm, 550 nm, 450 nm, and 750 nm, respectively. The field intensity distributions reveal that the maximum intensity values across all pixel channels are predominantly located at the corresponding pixel. This is particularly evident in the blue and near-infrared channels, although the red and green channels, due to the structural randomness, still present potential for further enhancement. To further verify the structure's sensitivity to variations in the incident angle and polarization angle of the light source, we conducted tests under different angular scenarios. The results indicated that changes in polarization angle had minimal impact, and the incident angle variation within range of -5 ° to +5 ° demonstrated excellent angular insensitivity (details can be found in **Figure S4**-**S6**). Thus, it is evident that the inverse-designed

RGB-IR routing structure maintains high optical efficiency while also exhibiting commendable angular insensitivity. This characteristic paves the way for future advancements in manufacturing and integration.

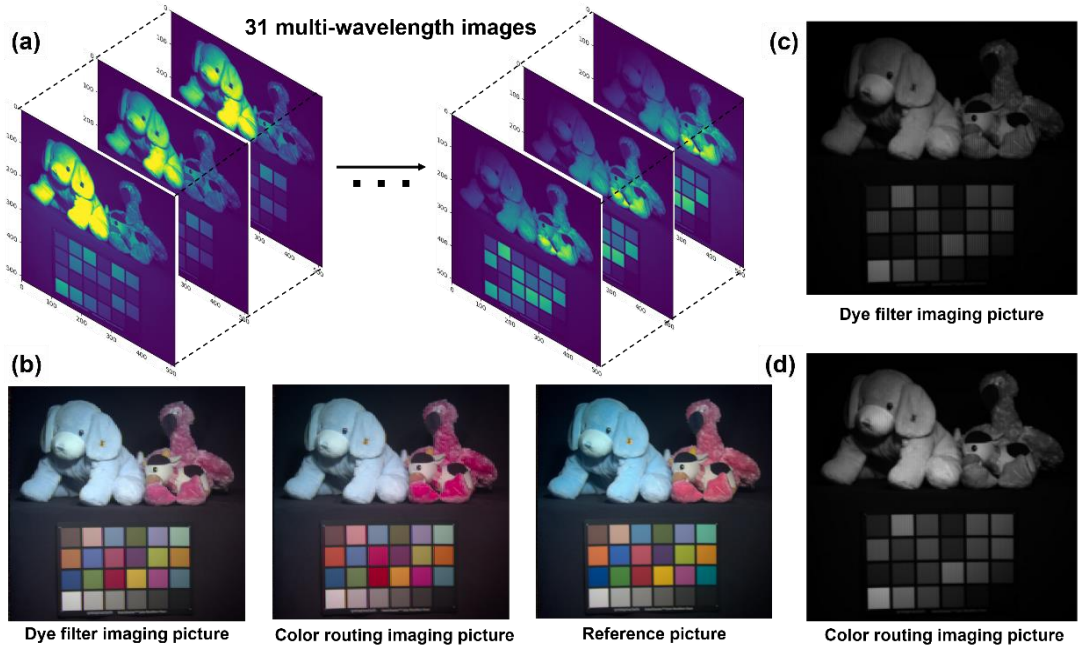

**Figure 4.** Multispectral image reconstruction using splitting spectra. (a) A dataset consisting of 31 multispectral images, with wavelengths ranging from 420 to 720 nm, and a bandwidth interval of 10 nm per image. (b) Comparison of color images simulated and reconstructed using color routing spectra with those using dye filter imaging, employing transformation matrix techniques. (c, d) Comparative analysis of grayscale signal intensities between images simulated using dye filter and color routing imaging.

After validating the spectral response of the inversely designed routing structure, we further investigated the imaging quality of the optimized routing structure by employing a conversion matrix method for imaging simulation.[44-46] A multispectral image, comprising 31 different waveband channels, was rendered using Python, and the corresponding multispectral images for each channel are illustrated in **Figure 4**a. In a multispectral image, each pixel corresponds to a specific incident value.[47,48] The combination of these incident values across the 31 channels yields the incident spectrum *L*. During the imaging simulation using the conversion matrix method, it is typically necessary to multiply the incident spectrum *L* of the image by the spectral response *S* of either the dye filter or the color routing structure. This multiplication produces the photocurrent vector *X* for each pixel. By subsequently multiplying the derived photocurrent vector *X* with the optimized conversion matrix, the photocurrent vector is transformed into an RGB vector, thereby enabling full-color imaging across multiple image channels.[49,50] To compare the imaging performance of the inverse-designed color routing structure with that of dye filters, **Figure 4**b presents the simulated imaging results of their respective spectral responses. It is evident that the image simulated using dye filters exhibits relatively muted colors, whereas the image generated through our inverse-designed color routing structure demonstrates more vivid colors, particularly in the red and near-infrared wavebands. This enhancement is closely linked to the high optical efficiency of the inverse-designed color routing structure. While comparing the color imaging quality, we also visualized the signal intensity differences between the dye filters and the inverse-designed color routing structure. This was achieved by integrating the obtained spectral response *S* with the incident spectra *L* of the various channels in the multispectral image, and then linearly converting the results into grayscale values, where a value of 255 corresponds to white color and the value of 0 corresponds to black color. The integrated light flux from the spectra is mapped to specific grayscale values, thereby producing simulated grayscale images that represent the imaging intensity. From the comparison between **Figure 4**c and **Figure 4**d, it can be observed that the simulated imaging effect of the color routing structure not only ensures color accuracy but also demonstrates higher signal intensity. This strategy, which achieves excellent imaging quality, holds significant promise as a potential solution for future image sensing technology.

**Conclusion**

In summary, our work successfully integrates the NSGA-II algorithm with the deep learning DCGAN model to achieve a Bayer-type Vis-NIR routing structure.

Compared to existing structural color techniques and dye filter-based image sensing strategies, we attained an optical efficiency enhancement exceeding 70% on a submicron pixel of 0.8 μm × 0.8 μm single-layer routing structure, with virtually no color distortion (shown in **Figure S7**). Furthermore, the inverse-designed color routing structure exhibits superior color imaging quality and grayscale signal intensity. This single-layer, inverse-designed color routing structure holds great promise for future application in on-chip image sensors through one-step lithography process. It is poised to play a significant role in high-resolution color imaging, near-infrared image sensing, and night vision navigation for autonomous driving, thereby offering substantial advancements for next-generation image sensor and photovoltaic technology.[51,52]

**Corresponding Author**

*E-mail: xianguang@jnu.edu.cn

**Notes**

The authors declare no competing financial interest

**Author ORCIDs**

Xianguang Yang, http://orcid.org/0000-0002-6787-9924

**Data Availability**

The data supporting this study are available from the corresponding author upon reasonable request.

**Acknowledgments**

This work was partially supported by the National Natural Science Foundation of China (11804120, 61827822, 62220106001 and 92050108), the Guangdong Basic and Applied Basic Research Foundation (2023A1515030209), the Key Research and Development Program of Guangdong Province (2023B0101200009), the Research

**TOC figure**

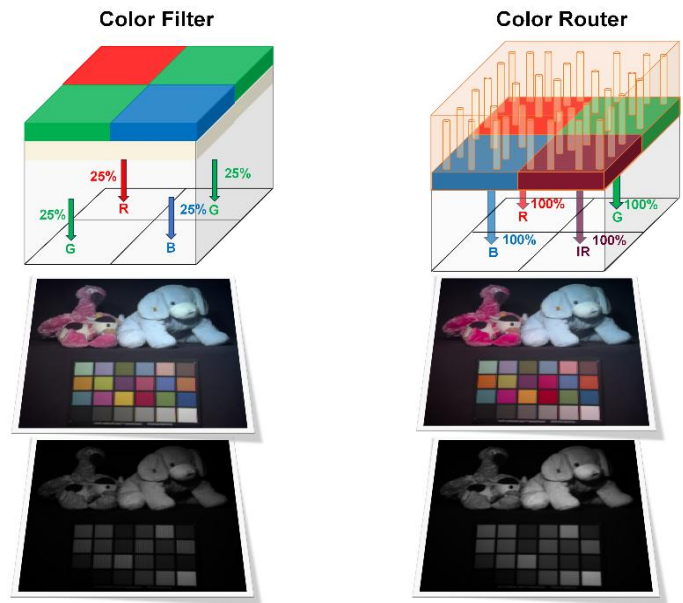

On-chip RGB-NIR routing realized by an inverse-designed single-layer meta-structure promotes over 70% signal enhancement with negligible color distortion and robust polarization insensitivity in image sensors with 800 nm pixel sizes.